%% file: cord_fer.tex
\documentclass{elsart}
\usepackage{natbib}
\usepackage{graphicx}
\begin{document}
\begin{frontmatter}
\title{Complex Entropic Forms for Gradient Pattern Analysis of Spatio-temporal
Dynamics}
\author[Inpe]{Fernando M. Ramos\thanksref{Fernando}}
\author[Inpe]{Reinaldo R. Rosa}
\author[Inpe]{Camilo Rodrigues Neto}
\author[Inpe]{Ademilson Zanandrea}

\address[Inpe]{LAC-INPE, Cx. Postal 515, 12201-970,S{\~a} Jo{\'e} dos Campos, Brazil}
\thanks[Fernando]{Corresponding author: F.M.R., fernando@lac.inpe.br}

\begin{abstract}

In this paper we 
describe two new computational operators, called {\it complex entropic form} (CEF)
and {\it generalized complex entropic form} (GEF), 
for pattern characterization of spatially extended systems. Besides of being 
a measure of regularity, both operators permit to quantify the degree of phase disorder
associated with a given gradient field. An application of CEF and GEF to the
analysis of the gradient pattern dynamics of a logistic Coupled Map Lattice is presented. 
Simulations using a Gaussian and random initial condition, provide interesting insights 
on the the system gradual transition from order/symmetry to disorder/randomness.

\end{abstract}


\end{frontmatter}

\section{Introduction} 
 
Nonlinear spatially extended dynamical systems yield complex amplitude patterns
which arise from the coupled dynamics of their different regions. 
Experiments in a variety 
of areas have shown spatio-temporal complexity,
notably in fluid flows, diffusive-reactive chemical systems, optical
electronics and laser physics. As reported by many authors (e.g. [1]), 
these systems cover a wide range of
scales, from millimeters in optical electronics experiments to thousands
of kilometers in natural systems. In
all this cases it is important to quantify the degree of local complexity
in order to characterize the spatial patterns and also study their
time evolution. Consequently, one of the main tasks in computational physics today is 
to define suitable measures to characterize the dynamics of such complex systems.
As stressed by Chat\`e [2], the current understanding 
of spatio-temporal disorder in extended systems is still very limited due to the lack of 
quantitative, meaningful, and sufficiently universal patterns and regimes 
characterizations.
 
Recently, the analysis of the gradient field of the amplitude envelope of dynamical 
extended systems has been shown to be a useful tool for understanding complex regimes as
intermittency and localized turbulence [3]. However, the classical measures of
complexity usually discharge the directional information contained in a vector field.
In this paper we describe two new computational operators, called {\it complex entropic 
form} (CEF) and {\it generalized complex entropic form} (GEF), 
for pattern characterization of spatially extended systems. Besides of being 
a measure of regularity, both operators permit to quantify the degree of phase disorder
associated with a given gradient field. Phase disorder plays an important role
in the analysis of pattern changing frequency in the amplitude domain [4].
For illustration purposes, we apply CEF and GEF to the
analysis of the gradient pattern dynamics of a logistic Coupled Map Lattice.

\section{Complex Entropic Form (CEF)}

If we consider spatially extended systems in two dimensions 
(x,y), their energy amplitude distribution is represented by the envelope A(x,y), 
which can be approximated by a matrix of amplitudes ${\bf A}=\{a_{k,l}\}$, 
with $K \times L$ pixels. Note that a dynamical sequence of matrices can be related to a 
temporal evolution of an envelope A(x,y,t). From the definition of ${\bf A}$,
it is possible to represent the gradient field of the amplitude envelope by
${\bf G} \equiv \nabla({\bf A})=\{z_{k,l}\}$, where $z_{k,l}$ is a complex number with 
$Re(z_{k,l})=a_{k,l+1} - a_{k,l-1}$ and $Im(z_{k,l})=a_{k+1,l} - a_{k-1,l}$.

Given the gradient field matrix $\nabla({\bf A})$, we now define the Complex Entropic 
Form (CEF) operator as:

\begin{equation}
\label{CEF}
S_c ({\bf G}) \equiv  - \sum_{k,l}  \frac{| z_{k,l} |}{| z |} 
\ln (\frac{z_{k,l}}{| z |}) = 
- \sum_{k,l}  \frac{| z_{k,l} |}{| z |} 
\ln (\frac{| z_{k,l} |}{| z |}) 
- i \sum_{k,l}  \frac{| z_{k,l} |}{| z |} \phi_{k,l}
\end{equation}
where $| z | = \sum_{k,l} | z_{k,l} |$, $| z_{k,l} |$ and $\phi_{k,l}$ are respectively
the modulus and the argument of $z_{k,l}$ and $i=\sqrt{-1}$.  

From the above definition, we immediately verify that the $Re(S_c)$ corresponds
to the classical Shannon's entropy measure $S$ of the moduli $|z_{k,l}|$ and that 
$Im(S_c)$ represents
the weighted average phase of the gradient field.
Consequently, a random generated pattern whose gradient field
displays no dominant direction, will have $Im(S_c) \approx 0$ and $Re(S_C)$
close to its maximum value, $Re(S_c)=log(M)$, where $M = K \times L$ is the number
of elements in the gradient matrix. On the other hand, intermittencies 
and symmetry breaking patterns lead to, respectively, a decrease
on the value of $Re(S_c)$ and to a non null $Im(S_c)$.
 
\section{Generalized Complex Entropic Form (GEF)} 

A more general entropic form that also incorporates the phase information
of the gradient field can be obtained from a generalization of the concept
of degeneracy $W$, given by the multinomial coefficient formula, and normally used to 
deduce the expression 
of Shannon's entropy of positive scalar fields, such as images [5]. 
Considering the gradient matrix ${\bf G}$ defined in the previous section, 
$W(z_{1,1},\ldots,z_{K,L})$ may be generalized as follows:

\begin{equation}
\label{W}
W(z_{1,1},\ldots,z_{K,L}) \equiv \frac{\Gamma(z)}{\Gamma(z_{1,1}) \ldots \Gamma(z_{K,L})}~~,
\end{equation}
where $z = \sum_{k,l} z_{k,l}$. Using Stirling's approximation, we immediately have

\begin{equation}
\label{GEF}
z^{-1} \ln W \longrightarrow S_z = - \sum_{k,l}  \frac{z_{k,l}}{z} 
\ln (\frac{z_{k,l}}{z})~~.
\end{equation}   
This expression, that we hereafter call Generalized Complex Entropic Form (GEF), displays
interesting properties [6], several of which it shares with Shannon's entropy. In particular,
we can easily verify that $S_z$ is invariant under rotation and scaling of the vector field,
and that $Im(S_z)=0$ and $Re(S_z)=S(|z_{k,l}|)$, when all $z_{k,l}$ have the same
direction (in other words, no information is conveyed by the phases when they are all
the same). 

\section{Application}

In order to illustrate the performance of the computational operators described
in the previous sections, we applied CEF and GEF to the 
characterization of amplitude pattern formation (fragmentation, symmetry breaking and 
phase disorder) of a logistic extended Coupled Map Lattice (CML). Additional results
of the performance of these operators in different contexts (osmosedimentation process [7],
wavelet spectra of temporal series [8], Swift-Hohenberg dynamics [9], porous silicon 
surface patterns [10]) are reported elsewhere.

CML is one of the most important systems that have been introduced to study the dynamics 
of spatio-temporal complexity in the amplitude domain [11]. The explicit form of a 
global CML is given by

\begin{equation}
\label{CML1}
x_{n+1} = (1 - \epsilon) f(x_n(k)) + \frac{\epsilon}{N} \sum_l^M f(x_n(l))~~,
\end{equation}                                                              
where $n$ refers to a  discrete time step and $M=K=L$ is the lattice size. Usually, 
the function  f(x) is chosen to be the well known dissipative chaotic logistic map,
$f(x) = 1 - \rho x^2$. 

From Eq. (\ref{CML1}), considering the matrix of amplitudes ${\bf A}=\{a_{k,l}\}$,
we have

\begin{equation}
\label{CML2}
a^{n+1}_{k,l} = (1 - \epsilon) a^n_{k,l} + \frac{\epsilon}{4} 
( f(a^n_{k+1,l}) + f(a^n_{k,l+1}) + f(a^n_{k-1,l}) + f(a^n_{k,l-1}) )~~,
\end{equation}
where $\epsilon$ is the coupling constant and $f(a) = \rho a(1-a)$, with $\rho$ being 
a real constant (the chaotic control parameter). 

We solved Eq. (\ref{CML2}), for $\epsilon = 0.5$, $\rho = 4.0$
(chaotic regime), and an aspect ratio (ratio between largest and smallest system 
characteristic scales) 
of $\Gamma =400$. We started the simulations from two different initial conditions:
(i) a Gaussian, symmetric initial distribution, and (ii) a random distribution. 
Figure 1 (a,b,c,d) shows four frames ($n=5,20,35,50$) of the evolution of the
gradient field and the corresponding intensity contours plots for the Gaussian
initial condition. Figures 2 and 3 depicts respectively the variation of $Re(S_z)$
and $Re(S_c)$, for test cases (i) and (ii). 

While the results of the random test case, as expected, do not display any 
particular characteristic, from the evolution of $Re(S_z)$ for the Gaussian  
initial distribution (Figure 2) we identify three well defined periods, corresponding 
to different phase 
disorder pattern regimes. For $n \leq 5$, the value of $Re(S_z)$ is very high 
(but fast decreasing), since the gradient field still presents a high degree of 
symmetry. For $5 \leq n \leq 20$, as the Gaussian perturbation reaches the boundaries
of the system, the symmetries of the initial condition are almost completely lost. 
Moreover, during this period, the system have not yet acquired the symmetries, in a 
{\it statistical} sense, that characterizes a random field and, thus, $Re(S_z)_G < Re(S_z)_R$. 
Finally, for $n > 20$, the gradient field becomes increasingly random. 
However, since the system still carries a memory of the original symmetry, 
the value of $Re(S_z)_G$ is higher than $Re(S_z)_R$ (see [12] for 
a similar analysis, concerning the loss and recovery of symmetries of Navier-Stokes equation 
for increasing values of Reynolds number).
The evolution of $Re(S_c)$ (Figure 3), although with less detail, also shows the system gradual 
transition from order/symmetry (lower $Re(S_c)$ value) to disorder/randomness
(higher $Re(S_c)$ value).  

\section{Concluding Remarks} 

In this paper we 
introduced two new computational operators, called {\it complex entropic form} (CEF)
and {\it generalized complex entropic form} (GEF), for pattern characterization of 
spatially extended systems. The real part of CEF corresponds
to the classical Shannon's entropy measure, while its imaginary part represents
the weighted average phase of the gradient field. The definition of GEF is
obtained from a generalization of the concept of degeneracy. Besides of being 
a measure of regularity or smoothness of the embedding amplitude envelope, both operators 
provide information on the degree of phase disorder of the corresponding gradient field. 
 
For illustration purposes, we applied CEF and GEF to the
analysis of the gradient pattern dynamics of a logistic Coupled Map Lattice. Simulations
using a Gaussian and random initial condition, provided interesting insights on the
the system gradual transition from order/symmetry to disorder/randomness.   
 
\section*{Acknowledgments} 
 
This work was supported by FAPESP-Brazil.
FMR also acknowledges the support given by CNPq-Brazil through
the research grant 300171/97-8.

\section*{References}
 
\begin{enumerate} 

\item H. L. Swinney, in {\it Time Series Prediction}, A. S. Weigend and N. A. 
Gershenfeld, eds., Addison-Wesley, Reading, 1994.

\item H. Chat\'e, {\it Physica D} {\bf 86} (1995) 238.

\item R. R. Rosa, A. S. Sharma, J. A. Valdivia, {\it Physica A} {\bf 257} (1998) 509.

\item R. R. Rosa, F. M. Ramos, C. Rodrigues Neto, J. A. Valdivia and A. S. Sharma,
{\it J. Braz. Soc. Mech. Sci.}, to appear.

\item B. R. Frieden, {\it Comp. Graph. Image Proc.} {bf 12} (1980) 40.

\item F. M. Ramos, C. Rodrigues Neto and R. R. Rosa, in preparation.

\item C. Rodrigues Neto, A. O. Fortuna, R. R. Rosa, F. M. Ramos, A. Zanandrea and 
M. A. Tenan, LAWNP'99, October 1999, Cordoba.

\item C. Rodrigues Neto, R. R. Rosa, F. M. Ramos and A. Zanandrea, SPIE'99, July 1999, 
Denver. 

\item R. R. Rosa, J. Pontes, C. I. Christov, D. Walgraef, F. M. Ramos and 
C. Rodrigues Neto, LAWNP'99, October 1999, Cordoba.

\item A. Ferreira da Silva, R. R. Rosa, F. M. Ramos, C. Rodrigues Neto, 
P. W. M. Machado, L. S. Roman and E. Veje, LAWNP'99, October 1999, 
Cordoba.

\item G. Perez, S. Sinha and H. A. Cerdeira, {\it Physica D} {\bf 63} (1993) 341.

\item U. Frisch, {\it Turbulence}, Cambridge University Press, Cambridge, 1995.

\end{enumerate}

\newpage
\section*{Figure Captions}

\bigskip
\parindent=0pt \textbf{Fig.\ 1} Four frames ($n=5, 20, 35, 50$) showing the evolution 
of the gradient field and intensity contours plots for the solution of logistic 
Coupled Map with Gaussian initial condition, $\Gamma =400$, $\epsilon= 0.5$ 
and  $\rho= 4.0$.
 
\bigskip
\parindent=0pt \textbf{Fig.\ 2} Evolution of $Re(S_z)$ as a function of the
time index $n$, for Gaussian and random initial conditions. 
 
\bigskip
\parindent=0pt \textbf{Fig.\ 3} Evolution of $Re(S_c)$ as a function of the
time index $n$, for Gaussian and random initial conditions.

\newpage

\begin{figure}[ccc]
\begin{center}
\includegraphics[width=13cm]{fig1.epsi}
\end{center}
\end{figure}

\clearpage

\begin{figure}[ttt]
\begin{center}
\input{f2}
\end{center}
\end{figure}

\begin{figure}[bbb]
\begin{center}
\input{f3}
\end{center}
\end{figure}

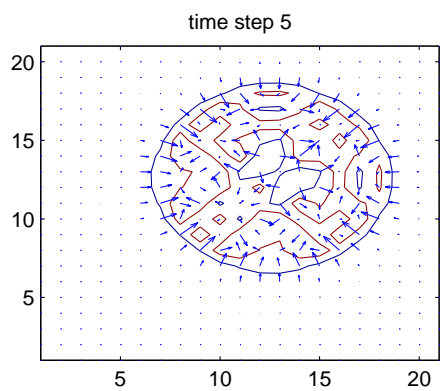
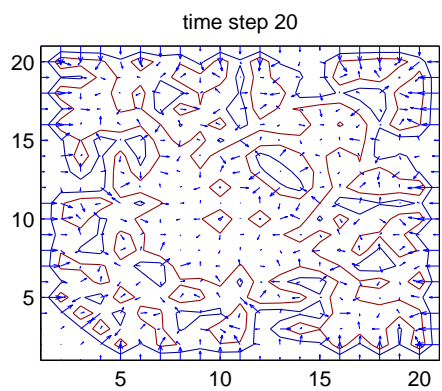
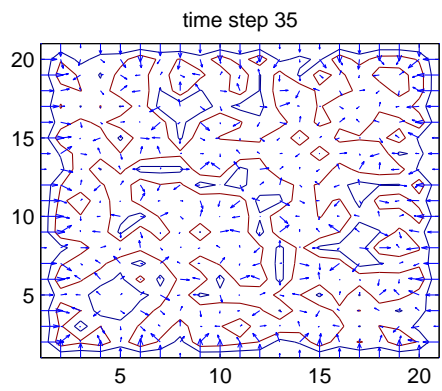
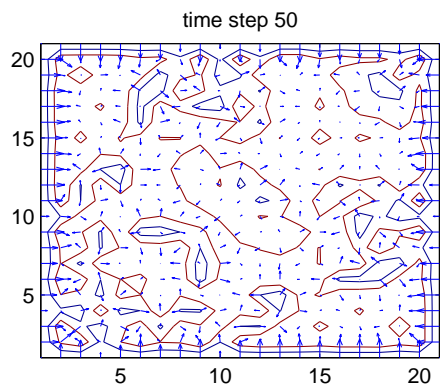
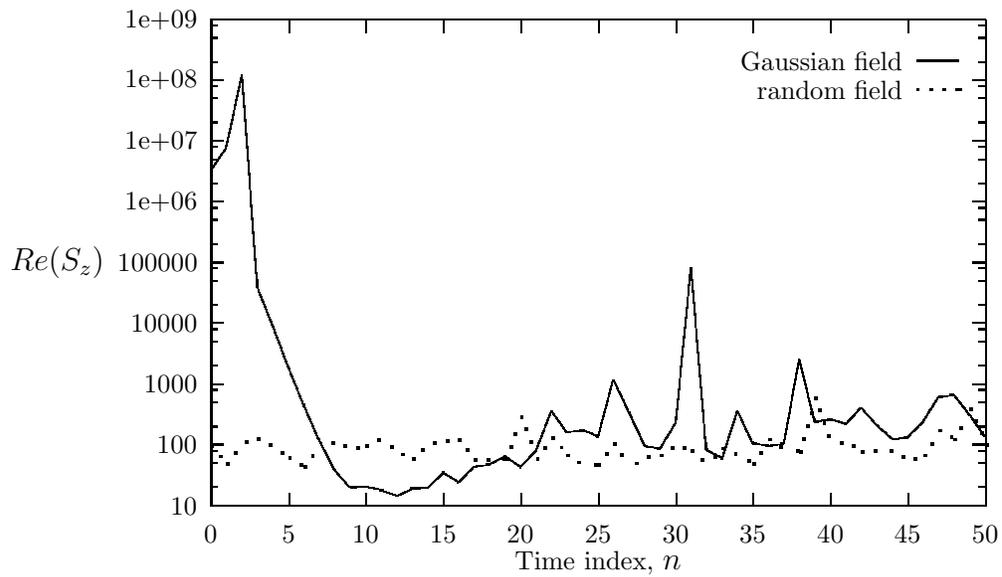
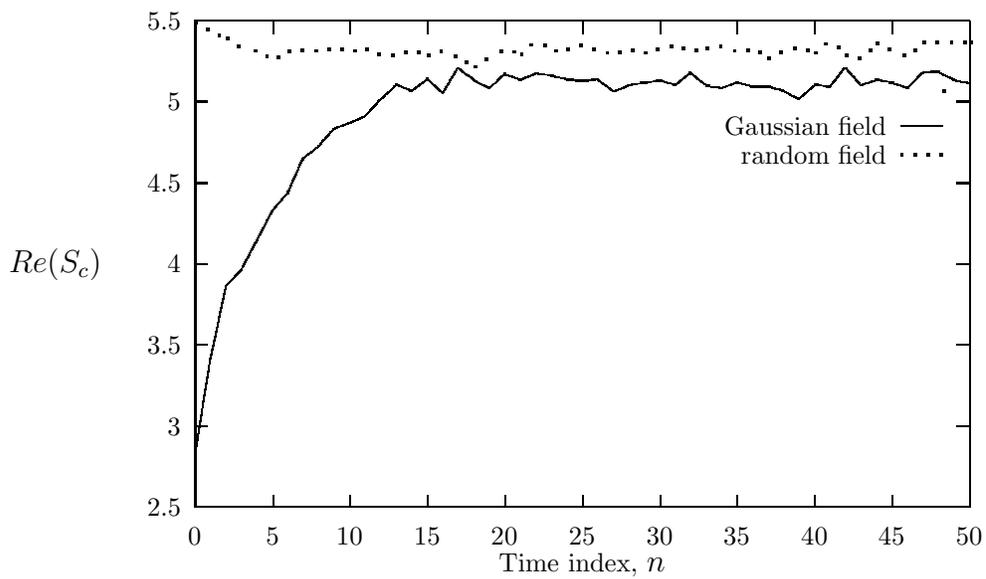
\end{document}

%% file: f2.tex
\setlength{\unitlength}{0.240900pt}
\ifx\plotpoint\undefined\newsavebox{\plotpoint}\fi
\sbox{\plotpoint}{\rule[-0.200pt]{0.400pt}{0.400pt}}%
\begin{picture}(1500,900)(0,0)
\font\gnuplot=cmr10 at 10pt
\gnuplot
\sbox{\plotpoint}{\rule[-0.200pt]{0.400pt}{0.400pt}}%
\put(220.0,113.0){\rule[-0.200pt]{0.400pt}{184.048pt}}
\put(220.0,113.0){\rule[-0.200pt]{4.818pt}{0.400pt}}
\put(198,113){\makebox(0,0)[r]{10}}
\put(1416.0,113.0){\rule[-0.200pt]{4.818pt}{0.400pt}}
\put(220.0,142.0){\rule[-0.200pt]{2.409pt}{0.400pt}}
\put(1426.0,142.0){\rule[-0.200pt]{2.409pt}{0.400pt}}
\put(220.0,180.0){\rule[-0.200pt]{2.409pt}{0.400pt}}
\put(1426.0,180.0){\rule[-0.200pt]{2.409pt}{0.400pt}}
\put(220.0,199.0){\rule[-0.200pt]{2.409pt}{0.400pt}}
\put(1426.0,199.0){\rule[-0.200pt]{2.409pt}{0.400pt}}
\put(220.0,209.0){\rule[-0.200pt]{4.818pt}{0.400pt}}
\put(198,209){\makebox(0,0)[r]{100}}
\put(1416.0,209.0){\rule[-0.200pt]{4.818pt}{0.400pt}}
\put(220.0,237.0){\rule[-0.200pt]{2.409pt}{0.400pt}}
\put(1426.0,237.0){\rule[-0.200pt]{2.409pt}{0.400pt}}
\put(220.0,275.0){\rule[-0.200pt]{2.409pt}{0.400pt}}
\put(1426.0,275.0){\rule[-0.200pt]{2.409pt}{0.400pt}}
\put(220.0,295.0){\rule[-0.200pt]{2.409pt}{0.400pt}}
\put(1426.0,295.0){\rule[-0.200pt]{2.409pt}{0.400pt}}
\put(220.0,304.0){\rule[-0.200pt]{4.818pt}{0.400pt}}
\put(198,304){\makebox(0,0)[r]{1000}}
\put(1416.0,304.0){\rule[-0.200pt]{4.818pt}{0.400pt}}
\put(220.0,333.0){\rule[-0.200pt]{2.409pt}{0.400pt}}
\put(1426.0,333.0){\rule[-0.200pt]{2.409pt}{0.400pt}}
\put(220.0,371.0){\rule[-0.200pt]{2.409pt}{0.400pt}}
\put(1426.0,371.0){\rule[-0.200pt]{2.409pt}{0.400pt}}
\put(220.0,390.0){\rule[-0.200pt]{2.409pt}{0.400pt}}
\put(1426.0,390.0){\rule[-0.200pt]{2.409pt}{0.400pt}}
\put(220.0,400.0){\rule[-0.200pt]{4.818pt}{0.400pt}}
\put(198,400){\makebox(0,0)[r]{10000}}
\put(1416.0,400.0){\rule[-0.200pt]{4.818pt}{0.400pt}}
\put(220.0,428.0){\rule[-0.200pt]{2.409pt}{0.400pt}}
\put(1426.0,428.0){\rule[-0.200pt]{2.409pt}{0.400pt}}
\put(220.0,466.0){\rule[-0.200pt]{2.409pt}{0.400pt}}
\put(1426.0,466.0){\rule[-0.200pt]{2.409pt}{0.400pt}}
\put(220.0,486.0){\rule[-0.200pt]{2.409pt}{0.400pt}}
\put(1426.0,486.0){\rule[-0.200pt]{2.409pt}{0.400pt}}
\put(220.0,495.0){\rule[-0.200pt]{4.818pt}{0.400pt}}
\put(198,495){\makebox(0,0)[r]{100000}}
\put(1416.0,495.0){\rule[-0.200pt]{4.818pt}{0.400pt}}
\put(220.0,524.0){\rule[-0.200pt]{2.409pt}{0.400pt}}
\put(1426.0,524.0){\rule[-0.200pt]{2.409pt}{0.400pt}}
\put(220.0,562.0){\rule[-0.200pt]{2.409pt}{0.400pt}}
\put(1426.0,562.0){\rule[-0.200pt]{2.409pt}{0.400pt}}
\put(220.0,581.0){\rule[-0.200pt]{2.409pt}{0.400pt}}
\put(1426.0,581.0){\rule[-0.200pt]{2.409pt}{0.400pt}}
\put(220.0,591.0){\rule[-0.200pt]{4.818pt}{0.400pt}}
\put(198,591){\makebox(0,0)[r]{1e+06}}
\put(1416.0,591.0){\rule[-0.200pt]{4.818pt}{0.400pt}}
\put(220.0,619.0){\rule[-0.200pt]{2.409pt}{0.400pt}}
\put(1426.0,619.0){\rule[-0.200pt]{2.409pt}{0.400pt}}
\put(220.0,657.0){\rule[-0.200pt]{2.409pt}{0.400pt}}
\put(1426.0,657.0){\rule[-0.200pt]{2.409pt}{0.400pt}}
\put(220.0,677.0){\rule[-0.200pt]{2.409pt}{0.400pt}}
\put(1426.0,677.0){\rule[-0.200pt]{2.409pt}{0.400pt}}
\put(220.0,686.0){\rule[-0.200pt]{4.818pt}{0.400pt}}
\put(198,686){\makebox(0,0)[r]{1e+07}}
\put(1416.0,686.0){\rule[-0.200pt]{4.818pt}{0.400pt}}
\put(220.0,715.0){\rule[-0.200pt]{2.409pt}{0.400pt}}
\put(1426.0,715.0){\rule[-0.200pt]{2.409pt}{0.400pt}}
\put(220.0,753.0){\rule[-0.200pt]{2.409pt}{0.400pt}}
\put(1426.0,753.0){\rule[-0.200pt]{2.409pt}{0.400pt}}
\put(220.0,772.0){\rule[-0.200pt]{2.409pt}{0.400pt}}
\put(1426.0,772.0){\rule[-0.200pt]{2.409pt}{0.400pt}}
\put(220.0,782.0){\rule[-0.200pt]{4.818pt}{0.400pt}}
\put(198,782){\makebox(0,0)[r]{1e+08}}
\put(1416.0,782.0){\rule[-0.200pt]{4.818pt}{0.400pt}}
\put(220.0,810.0){\rule[-0.200pt]{2.409pt}{0.400pt}}
\put(1426.0,810.0){\rule[-0.200pt]{2.409pt}{0.400pt}}
\put(220.0,848.0){\rule[-0.200pt]{2.409pt}{0.400pt}}
\put(1426.0,848.0){\rule[-0.200pt]{2.409pt}{0.400pt}}
\put(220.0,868.0){\rule[-0.200pt]{2.409pt}{0.400pt}}
\put(1426.0,868.0){\rule[-0.200pt]{2.409pt}{0.400pt}}
\put(220.0,877.0){\rule[-0.200pt]{4.818pt}{0.400pt}}
\put(198,877){\makebox(0,0)[r]{1e+09}}
\put(1416.0,877.0){\rule[-0.200pt]{4.818pt}{0.400pt}}
\put(220.0,113.0){\rule[-0.200pt]{0.400pt}{4.818pt}}
\put(220,68){\makebox(0,0){0}}
\put(220.0,857.0){\rule[-0.200pt]{0.400pt}{4.818pt}}
\put(342.0,113.0){\rule[-0.200pt]{0.400pt}{4.818pt}}
\put(342,68){\makebox(0,0){5}}
\put(342.0,857.0){\rule[-0.200pt]{0.400pt}{4.818pt}}
\put(463.0,113.0){\rule[-0.200pt]{0.400pt}{4.818pt}}
\put(463,68){\makebox(0,0){10}}
\put(463.0,857.0){\rule[-0.200pt]{0.400pt}{4.818pt}}
\put(585.0,113.0){\rule[-0.200pt]{0.400pt}{4.818pt}}
\put(585,68){\makebox(0,0){15}}
\put(585.0,857.0){\rule[-0.200pt]{0.400pt}{4.818pt}}
\put(706.0,113.0){\rule[-0.200pt]{0.400pt}{4.818pt}}
\put(706,68){\makebox(0,0){20}}
\put(706.0,857.0){\rule[-0.200pt]{0.400pt}{4.818pt}}
\put(828.0,113.0){\rule[-0.200pt]{0.400pt}{4.818pt}}
\put(828,68){\makebox(0,0){25}}
\put(828.0,857.0){\rule[-0.200pt]{0.400pt}{4.818pt}}
\put(950.0,113.0){\rule[-0.200pt]{0.400pt}{4.818pt}}
\put(950,68){\makebox(0,0){30}}
\put(950.0,857.0){\rule[-0.200pt]{0.400pt}{4.818pt}}
\put(1071.0,113.0){\rule[-0.200pt]{0.400pt}{4.818pt}}
\put(1071,68){\makebox(0,0){35}}
\put(1071.0,857.0){\rule[-0.200pt]{0.400pt}{4.818pt}}
\put(1193.0,113.0){\rule[-0.200pt]{0.400pt}{4.818pt}}
\put(1193,68){\makebox(0,0){40}}
\put(1193.0,857.0){\rule[-0.200pt]{0.400pt}{4.818pt}}
\put(1314.0,113.0){\rule[-0.200pt]{0.400pt}{4.818pt}}
\put(1314,68){\makebox(0,0){45}}
\put(1314.0,857.0){\rule[-0.200pt]{0.400pt}{4.818pt}}
\put(1436.0,113.0){\rule[-0.200pt]{0.400pt}{4.818pt}}
\put(1436,68){\makebox(0,0){50}}
\put(1436.0,857.0){\rule[-0.200pt]{0.400pt}{4.818pt}}
\put(220.0,113.0){\rule[-0.200pt]{292.934pt}{0.400pt}}
\put(1436.0,113.0){\rule[-0.200pt]{0.400pt}{184.048pt}}
\put(220.0,877.0){\rule[-0.200pt]{292.934pt}{0.400pt}}
\put(-21,495){\makebox(0,0){$Re(S_z)$}}
\put(828,23){\makebox(0,0){Time index, $n$}}
\put(220.0,113.0){\rule[-0.200pt]{0.400pt}{184.048pt}}
\put(1306,812){\makebox(0,0)[r]{Gaussian field}}
\put(1328.0,812.0){\rule[-0.200pt]{15.899pt}{0.400pt}}
\put(220,640){\usebox{\plotpoint}}
\multiput(220.58,640.00)(0.496,0.731){45}{\rule{0.120pt}{0.683pt}}
\multiput(219.17,640.00)(24.000,33.582){2}{\rule{0.400pt}{0.342pt}}
\multiput(244.58,675.00)(0.497,2.325){47}{\rule{0.120pt}{1.940pt}}
\multiput(243.17,675.00)(25.000,110.973){2}{\rule{0.400pt}{0.970pt}}
\multiput(269.58,766.48)(0.496,-7.056){45}{\rule{0.120pt}{5.667pt}}
\multiput(268.17,778.24)(24.000,-322.239){2}{\rule{0.400pt}{2.833pt}}
\multiput(293.58,451.43)(0.496,-1.260){45}{\rule{0.120pt}{1.100pt}}
\multiput(292.17,453.72)(24.000,-57.717){2}{\rule{0.400pt}{0.550pt}}
\multiput(317.58,391.20)(0.497,-1.330){47}{\rule{0.120pt}{1.156pt}}
\multiput(316.17,393.60)(25.000,-63.601){2}{\rule{0.400pt}{0.578pt}}
\multiput(342.58,325.50)(0.496,-1.239){45}{\rule{0.120pt}{1.083pt}}
\multiput(341.17,327.75)(24.000,-56.751){2}{\rule{0.400pt}{0.542pt}}
\multiput(366.58,266.78)(0.496,-1.154){45}{\rule{0.120pt}{1.017pt}}
\multiput(365.17,268.89)(24.000,-52.890){2}{\rule{0.400pt}{0.508pt}}
\multiput(390.58,212.40)(0.497,-0.965){47}{\rule{0.120pt}{0.868pt}}
\multiput(389.17,214.20)(25.000,-46.198){2}{\rule{0.400pt}{0.434pt}}
\multiput(415.58,165.72)(0.496,-0.562){45}{\rule{0.120pt}{0.550pt}}
\multiput(414.17,166.86)(24.000,-25.858){2}{\rule{0.400pt}{0.275pt}}
\put(439,141.17){\rule{4.900pt}{0.400pt}}
\multiput(439.00,140.17)(13.830,2.000){2}{\rule{2.450pt}{0.400pt}}
\multiput(463.00,141.93)(2.208,-0.482){9}{\rule{1.767pt}{0.116pt}}
\multiput(463.00,142.17)(21.333,-6.000){2}{\rule{0.883pt}{0.400pt}}
\multiput(488.00,135.93)(1.368,-0.489){15}{\rule{1.167pt}{0.118pt}}
\multiput(488.00,136.17)(21.579,-9.000){2}{\rule{0.583pt}{0.400pt}}
\multiput(512.00,128.58)(1.013,0.492){21}{\rule{0.900pt}{0.119pt}}
\multiput(512.00,127.17)(22.132,12.000){2}{\rule{0.450pt}{0.400pt}}
\put(536,139.67){\rule{5.782pt}{0.400pt}}
\multiput(536.00,139.17)(12.000,1.000){2}{\rule{2.891pt}{0.400pt}}
\multiput(560.00,141.58)(0.519,0.496){45}{\rule{0.517pt}{0.120pt}}
\multiput(560.00,140.17)(23.928,24.000){2}{\rule{0.258pt}{0.400pt}}
\multiput(585.00,163.92)(0.753,-0.494){29}{\rule{0.700pt}{0.119pt}}
\multiput(585.00,164.17)(22.547,-16.000){2}{\rule{0.350pt}{0.400pt}}
\multiput(609.58,149.00)(0.496,0.519){45}{\rule{0.120pt}{0.517pt}}
\multiput(608.17,149.00)(24.000,23.928){2}{\rule{0.400pt}{0.258pt}}
\multiput(633.00,174.60)(3.552,0.468){5}{\rule{2.600pt}{0.113pt}}
\multiput(633.00,173.17)(19.604,4.000){2}{\rule{1.300pt}{0.400pt}}
\multiput(658.00,178.58)(0.933,0.493){23}{\rule{0.838pt}{0.119pt}}
\multiput(658.00,177.17)(22.260,13.000){2}{\rule{0.419pt}{0.400pt}}
\multiput(682.00,189.92)(0.668,-0.495){33}{\rule{0.633pt}{0.119pt}}
\multiput(682.00,190.17)(22.685,-18.000){2}{\rule{0.317pt}{0.400pt}}
\multiput(706.58,173.00)(0.497,0.539){47}{\rule{0.120pt}{0.532pt}}
\multiput(705.17,173.00)(25.000,25.896){2}{\rule{0.400pt}{0.266pt}}
\multiput(731.58,200.00)(0.496,1.302){45}{\rule{0.120pt}{1.133pt}}
\multiput(730.17,200.00)(24.000,59.648){2}{\rule{0.400pt}{0.567pt}}
\multiput(755.58,259.23)(0.496,-0.710){45}{\rule{0.120pt}{0.667pt}}
\multiput(754.17,260.62)(24.000,-32.616){2}{\rule{0.400pt}{0.333pt}}
\multiput(779.00,228.60)(3.552,0.468){5}{\rule{2.600pt}{0.113pt}}
\multiput(779.00,227.17)(19.604,4.000){2}{\rule{1.300pt}{0.400pt}}
\multiput(804.00,230.92)(1.109,-0.492){19}{\rule{0.973pt}{0.118pt}}
\multiput(804.00,231.17)(21.981,-11.000){2}{\rule{0.486pt}{0.400pt}}
\multiput(828.58,221.00)(0.496,1.895){45}{\rule{0.120pt}{1.600pt}}
\multiput(827.17,221.00)(24.000,86.679){2}{\rule{0.400pt}{0.800pt}}
\multiput(852.58,307.13)(0.497,-1.046){47}{\rule{0.120pt}{0.932pt}}
\multiput(851.17,309.07)(25.000,-50.066){2}{\rule{0.400pt}{0.466pt}}
\multiput(877.58,254.99)(0.496,-1.091){45}{\rule{0.120pt}{0.967pt}}
\multiput(876.17,256.99)(24.000,-49.994){2}{\rule{0.400pt}{0.483pt}}
\multiput(901.00,205.93)(2.602,-0.477){7}{\rule{2.020pt}{0.115pt}}
\multiput(901.00,206.17)(19.807,-5.000){2}{\rule{1.010pt}{0.400pt}}
\multiput(925.58,202.00)(0.497,0.823){47}{\rule{0.120pt}{0.756pt}}
\multiput(924.17,202.00)(25.000,39.431){2}{\rule{0.400pt}{0.378pt}}
\multiput(950.58,243.00)(0.496,5.152){45}{\rule{0.120pt}{4.167pt}}
\multiput(949.17,243.00)(24.000,235.352){2}{\rule{0.400pt}{2.083pt}}
\multiput(974.58,466.80)(0.496,-6.041){45}{\rule{0.120pt}{4.867pt}}
\multiput(973.17,476.90)(24.000,-275.899){2}{\rule{0.400pt}{2.433pt}}
\multiput(998.00,199.92)(0.901,-0.494){25}{\rule{0.814pt}{0.119pt}}
\multiput(998.00,200.17)(23.310,-14.000){2}{\rule{0.407pt}{0.400pt}}
\multiput(1023.58,187.00)(0.496,1.577){45}{\rule{0.120pt}{1.350pt}}
\multiput(1022.17,187.00)(24.000,72.198){2}{\rule{0.400pt}{0.675pt}}
\multiput(1047.58,258.06)(0.496,-1.069){45}{\rule{0.120pt}{0.950pt}}
\multiput(1046.17,260.03)(24.000,-49.028){2}{\rule{0.400pt}{0.475pt}}
\multiput(1071.00,209.94)(3.552,-0.468){5}{\rule{2.600pt}{0.113pt}}
\multiput(1071.00,210.17)(19.604,-4.000){2}{\rule{1.300pt}{0.400pt}}
\multiput(1096.00,207.60)(3.406,0.468){5}{\rule{2.500pt}{0.113pt}}
\multiput(1096.00,206.17)(18.811,4.000){2}{\rule{1.250pt}{0.400pt}}
\multiput(1120.58,211.00)(0.496,2.783){45}{\rule{0.120pt}{2.300pt}}
\multiput(1119.17,211.00)(24.000,127.226){2}{\rule{0.400pt}{1.150pt}}
\multiput(1144.58,335.80)(0.496,-2.064){45}{\rule{0.120pt}{1.733pt}}
\multiput(1143.17,339.40)(24.000,-94.402){2}{\rule{0.400pt}{0.867pt}}
\multiput(1168.00,245.60)(3.552,0.468){5}{\rule{2.600pt}{0.113pt}}
\multiput(1168.00,244.17)(19.604,4.000){2}{\rule{1.300pt}{0.400pt}}
\multiput(1193.00,247.93)(1.550,-0.488){13}{\rule{1.300pt}{0.117pt}}
\multiput(1193.00,248.17)(21.302,-8.000){2}{\rule{0.650pt}{0.400pt}}
\multiput(1217.58,241.00)(0.496,0.541){45}{\rule{0.120pt}{0.533pt}}
\multiput(1216.17,241.00)(24.000,24.893){2}{\rule{0.400pt}{0.267pt}}
\multiput(1241.58,264.73)(0.497,-0.559){47}{\rule{0.120pt}{0.548pt}}
\multiput(1240.17,265.86)(25.000,-26.863){2}{\rule{0.400pt}{0.274pt}}
\multiput(1266.00,237.92)(0.544,-0.496){41}{\rule{0.536pt}{0.120pt}}
\multiput(1266.00,238.17)(22.887,-22.000){2}{\rule{0.268pt}{0.400pt}}
\multiput(1290.00,217.60)(3.406,0.468){5}{\rule{2.500pt}{0.113pt}}
\multiput(1290.00,216.17)(18.811,4.000){2}{\rule{1.250pt}{0.400pt}}
\multiput(1314.00,221.58)(0.498,0.497){47}{\rule{0.500pt}{0.120pt}}
\multiput(1314.00,220.17)(23.962,25.000){2}{\rule{0.250pt}{0.400pt}}
\multiput(1339.58,246.00)(0.496,0.794){45}{\rule{0.120pt}{0.733pt}}
\multiput(1338.17,246.00)(24.000,36.478){2}{\rule{0.400pt}{0.367pt}}
\multiput(1363.00,284.61)(5.151,0.447){3}{\rule{3.300pt}{0.108pt}}
\multiput(1363.00,283.17)(17.151,3.000){2}{\rule{1.650pt}{0.400pt}}
\multiput(1387.58,284.46)(0.497,-0.640){47}{\rule{0.120pt}{0.612pt}}
\multiput(1386.17,285.73)(25.000,-30.730){2}{\rule{0.400pt}{0.306pt}}
\multiput(1412.58,252.23)(0.496,-0.710){45}{\rule{0.120pt}{0.667pt}}
\multiput(1411.17,253.62)(24.000,-32.616){2}{\rule{0.400pt}{0.333pt}}
\sbox{\plotpoint}{\rule[-0.500pt]{1.000pt}{1.000pt}}%
\put(1306,767){\makebox(0,0)[r]{random field}}
\multiput(1328,767)(20.756,0.000){4}{\usebox{\plotpoint}}
\put(1394,767){\usebox{\plotpoint}}
\put(220,206){\usebox{\plotpoint}}
\multiput(220,206)(13.789,-15.513){2}{\usebox{\plotpoint}}
\multiput(244,179)(12.533,16.544){2}{\usebox{\plotpoint}}
\put(274.33,213.33){\usebox{\plotpoint}}
\multiput(293,218)(18.250,-9.885){2}{\usebox{\plotpoint}}
\put(330.01,196.15){\usebox{\plotpoint}}
\multiput(342,188)(17.928,-10.458){2}{\usebox{\plotpoint}}
\multiput(366,174)(11.969,16.957){2}{\usebox{\plotpoint}}
\put(409.62,211.14){\usebox{\plotpoint}}
\put(429.13,206.11){\usebox{\plotpoint}}
\put(448.85,204.05){\usebox{\plotpoint}}
\multiput(463,207)(18.712,8.982){2}{\usebox{\plotpoint}}
\put(502.81,205.43){\usebox{\plotpoint}}
\put(519.53,193.55){\usebox{\plotpoint}}
\multiput(536,186)(14.374,14.973){2}{\usebox{\plotpoint}}
\put(569.46,211.76){\usebox{\plotpoint}}
\put(590.13,213.64){\usebox{\plotpoint}}
\multiput(609,216)(12.706,-16.412){2}{\usebox{\plotpoint}}
\multiput(633,185)(20.739,-0.830){2}{\usebox{\plotpoint}}
\put(678.47,185.71){\usebox{\plotpoint}}
\multiput(682,186)(7.093,19.506){3}{\usebox{\plotpoint}}
\multiput(706,252)(7.451,-19.372){3}{\usebox{\plotpoint}}
\multiput(731,187)(12.208,16.786){2}{\usebox{\plotpoint}}
\multiput(755,220)(13.508,-15.759){2}{\usebox{\plotpoint}}
\multiput(779,192)(17.482,-11.188){2}{\usebox{\plotpoint}}
\put(823.02,176.79){\usebox{\plotpoint}}
\multiput(828,177)(12.208,16.786){2}{\usebox{\plotpoint}}
\multiput(852,210)(11.408,-17.340){2}{\usebox{\plotpoint}}
\put(887.21,179.23){\usebox{\plotpoint}}
\put(904.82,189.48){\usebox{\plotpoint}}
\multiput(925,192)(17.482,11.188){2}{\usebox{\plotpoint}}
\put(961.47,203.70){\usebox{\plotpoint}}
\multiput(974,199)(15.620,-13.668){2}{\usebox{\plotpoint}}
\put(1010.02,190.02){\usebox{\plotpoint}}
\multiput(1023,203)(19.159,-7.983){2}{\usebox{\plotpoint}}
\put(1062.47,183.98){\usebox{\plotpoint}}
\multiput(1071,179)(11.408,17.340){2}{\usebox{\plotpoint}}
\multiput(1096,217)(16.937,-11.997){2}{\usebox{\plotpoint}}
\put(1138.93,198.42){\usebox{\plotpoint}}
\multiput(1144,198)(5.702,19.957){4}{\usebox{\plotpoint}}
\multiput(1168,282)(6.643,-19.664){4}{\usebox{\plotpoint}}
\put(1208.91,211.31){\usebox{\plotpoint}}
\put(1227.39,206.07){\usebox{\plotpoint}}
\multiput(1241,197)(20.608,2.473){2}{\usebox{\plotpoint}}
\put(1286.71,199.14){\usebox{\plotpoint}}
\put(1304.81,189.74){\usebox{\plotpoint}}
\put(1323.45,187.02){\usebox{\plotpoint}}
\multiput(1339,192)(10.878,17.677){2}{\usebox{\plotpoint}}
\multiput(1363,231)(17.928,-10.458){2}{\usebox{\plotpoint}}
\multiput(1387,217)(9.588,18.408){2}{\usebox{\plotpoint}}
\multiput(1412,265)(8.176,-19.077){3}{\usebox{\plotpoint}}
\put(1436,209){\usebox{\plotpoint}}
\end{picture}

%% file: f3.tex
\setlength{\unitlength}{0.240900pt}
\ifx\plotpoint\undefined\newsavebox{\plotpoint}\fi
\sbox{\plotpoint}{\rule[-0.200pt]{0.400pt}{0.400pt}}%
\begin{picture}(1500,900)(0,0)
\font\gnuplot=cmr10 at 10pt
\gnuplot
\sbox{\plotpoint}{\rule[-0.200pt]{0.400pt}{0.400pt}}%
\put(220.0,113.0){\rule[-0.200pt]{0.400pt}{184.048pt}}
\put(220.0,113.0){\rule[-0.200pt]{4.818pt}{0.400pt}}
\put(198,113){\makebox(0,0)[r]{2.5}}
\put(1416.0,113.0){\rule[-0.200pt]{4.818pt}{0.400pt}}
\put(220.0,240.0){\rule[-0.200pt]{4.818pt}{0.400pt}}
\put(198,240){\makebox(0,0)[r]{3}}
\put(1416.0,240.0){\rule[-0.200pt]{4.818pt}{0.400pt}}
\put(220.0,368.0){\rule[-0.200pt]{4.818pt}{0.400pt}}
\put(198,368){\makebox(0,0)[r]{3.5}}
\put(1416.0,368.0){\rule[-0.200pt]{4.818pt}{0.400pt}}
\put(220.0,495.0){\rule[-0.200pt]{4.818pt}{0.400pt}}
\put(198,495){\makebox(0,0)[r]{4}}
\put(1416.0,495.0){\rule[-0.200pt]{4.818pt}{0.400pt}}
\put(220.0,622.0){\rule[-0.200pt]{4.818pt}{0.400pt}}
\put(198,622){\makebox(0,0)[r]{4.5}}
\put(1416.0,622.0){\rule[-0.200pt]{4.818pt}{0.400pt}}
\put(220.0,750.0){\rule[-0.200pt]{4.818pt}{0.400pt}}
\put(198,750){\makebox(0,0)[r]{5}}
\put(1416.0,750.0){\rule[-0.200pt]{4.818pt}{0.400pt}}
\put(220.0,877.0){\rule[-0.200pt]{4.818pt}{0.400pt}}
\put(198,877){\makebox(0,0)[r]{5.5}}
\put(1416.0,877.0){\rule[-0.200pt]{4.818pt}{0.400pt}}
\put(220.0,113.0){\rule[-0.200pt]{0.400pt}{4.818pt}}
\put(220,68){\makebox(0,0){0}}
\put(220.0,857.0){\rule[-0.200pt]{0.400pt}{4.818pt}}
\put(342.0,113.0){\rule[-0.200pt]{0.400pt}{4.818pt}}
\put(342,68){\makebox(0,0){5}}
\put(342.0,857.0){\rule[-0.200pt]{0.400pt}{4.818pt}}
\put(463.0,113.0){\rule[-0.200pt]{0.400pt}{4.818pt}}
\put(463,68){\makebox(0,0){10}}
\put(463.0,857.0){\rule[-0.200pt]{0.400pt}{4.818pt}}
\put(585.0,113.0){\rule[-0.200pt]{0.400pt}{4.818pt}}
\put(585,68){\makebox(0,0){15}}
\put(585.0,857.0){\rule[-0.200pt]{0.400pt}{4.818pt}}
\put(706.0,113.0){\rule[-0.200pt]{0.400pt}{4.818pt}}
\put(706,68){\makebox(0,0){20}}
\put(706.0,857.0){\rule[-0.200pt]{0.400pt}{4.818pt}}
\put(828.0,113.0){\rule[-0.200pt]{0.400pt}{4.818pt}}
\put(828,68){\makebox(0,0){25}}
\put(828.0,857.0){\rule[-0.200pt]{0.400pt}{4.818pt}}
\put(950.0,113.0){\rule[-0.200pt]{0.400pt}{4.818pt}}
\put(950,68){\makebox(0,0){30}}
\put(950.0,857.0){\rule[-0.200pt]{0.400pt}{4.818pt}}
\put(1071.0,113.0){\rule[-0.200pt]{0.400pt}{4.818pt}}
\put(1071,68){\makebox(0,0){35}}
\put(1071.0,857.0){\rule[-0.200pt]{0.400pt}{4.818pt}}
\put(1193.0,113.0){\rule[-0.200pt]{0.400pt}{4.818pt}}
\put(1193,68){\makebox(0,0){40}}
\put(1193.0,857.0){\rule[-0.200pt]{0.400pt}{4.818pt}}
\put(1314.0,113.0){\rule[-0.200pt]{0.400pt}{4.818pt}}
\put(1314,68){\makebox(0,0){45}}
\put(1314.0,857.0){\rule[-0.200pt]{0.400pt}{4.818pt}}
\put(1436.0,113.0){\rule[-0.200pt]{0.400pt}{4.818pt}}
\put(1436,68){\makebox(0,0){50}}
\put(1436.0,857.0){\rule[-0.200pt]{0.400pt}{4.818pt}}
\put(220.0,113.0){\rule[-0.200pt]{292.934pt}{0.400pt}}
\put(1436.0,113.0){\rule[-0.200pt]{0.400pt}{184.048pt}}
\put(220.0,877.0){\rule[-0.200pt]{292.934pt}{0.400pt}}
\put(1,495){\makebox(0,0){$Re(S_c)$}}
\put(828,23){\makebox(0,0){Time index, $n$}}
\put(220.0,113.0){\rule[-0.200pt]{0.400pt}{184.048pt}}
\put(1306,712){\makebox(0,0)[r]{Gaussian field}}
\put(1328.0,712.0){\rule[-0.200pt]{15.899pt}{0.400pt}}
\put(220,193){\usebox{\plotpoint}}
\multiput(220.58,193.00)(0.496,3.164){45}{\rule{0.120pt}{2.600pt}}
\multiput(219.17,193.00)(24.000,144.604){2}{\rule{0.400pt}{1.300pt}}
\multiput(244.58,343.00)(0.497,2.366){47}{\rule{0.120pt}{1.972pt}}
\multiput(243.17,343.00)(25.000,112.907){2}{\rule{0.400pt}{0.986pt}}
\multiput(269.58,460.00)(0.496,0.541){45}{\rule{0.120pt}{0.533pt}}
\multiput(268.17,460.00)(24.000,24.893){2}{\rule{0.400pt}{0.267pt}}
\multiput(293.58,486.00)(0.496,0.943){45}{\rule{0.120pt}{0.850pt}}
\multiput(292.17,486.00)(24.000,43.236){2}{\rule{0.400pt}{0.425pt}}
\multiput(317.58,531.00)(0.497,0.965){47}{\rule{0.120pt}{0.868pt}}
\multiput(316.17,531.00)(25.000,46.198){2}{\rule{0.400pt}{0.434pt}}
\multiput(342.58,579.00)(0.496,0.583){45}{\rule{0.120pt}{0.567pt}}
\multiput(341.17,579.00)(24.000,26.824){2}{\rule{0.400pt}{0.283pt}}
\multiput(366.58,607.00)(0.496,1.133){45}{\rule{0.120pt}{1.000pt}}
\multiput(365.17,607.00)(24.000,51.924){2}{\rule{0.400pt}{0.500pt}}
\multiput(390.00,661.58)(0.625,0.496){37}{\rule{0.600pt}{0.119pt}}
\multiput(390.00,660.17)(23.755,20.000){2}{\rule{0.300pt}{0.400pt}}
\multiput(415.58,681.00)(0.496,0.562){45}{\rule{0.120pt}{0.550pt}}
\multiput(414.17,681.00)(24.000,25.858){2}{\rule{0.400pt}{0.275pt}}
\multiput(439.00,708.59)(1.368,0.489){15}{\rule{1.167pt}{0.118pt}}
\multiput(439.00,707.17)(21.579,9.000){2}{\rule{0.583pt}{0.400pt}}
\multiput(463.00,717.58)(1.156,0.492){19}{\rule{1.009pt}{0.118pt}}
\multiput(463.00,716.17)(22.906,11.000){2}{\rule{0.505pt}{0.400pt}}
\multiput(488.58,728.00)(0.496,0.541){45}{\rule{0.120pt}{0.533pt}}
\multiput(487.17,728.00)(24.000,24.893){2}{\rule{0.400pt}{0.267pt}}
\multiput(512.00,754.58)(0.520,0.496){43}{\rule{0.517pt}{0.120pt}}
\multiput(512.00,753.17)(22.926,23.000){2}{\rule{0.259pt}{0.400pt}}
\multiput(536.00,775.92)(1.225,-0.491){17}{\rule{1.060pt}{0.118pt}}
\multiput(536.00,776.17)(21.800,-10.000){2}{\rule{0.530pt}{0.400pt}}
\multiput(560.00,767.58)(0.659,0.495){35}{\rule{0.626pt}{0.119pt}}
\multiput(560.00,766.17)(23.700,19.000){2}{\rule{0.313pt}{0.400pt}}
\multiput(585.00,784.92)(0.520,-0.496){43}{\rule{0.517pt}{0.120pt}}
\multiput(585.00,785.17)(22.926,-23.000){2}{\rule{0.259pt}{0.400pt}}
\multiput(609.58,763.00)(0.496,0.837){45}{\rule{0.120pt}{0.767pt}}
\multiput(608.17,763.00)(24.000,38.409){2}{\rule{0.400pt}{0.383pt}}
\multiput(633.00,801.92)(0.659,-0.495){35}{\rule{0.626pt}{0.119pt}}
\multiput(633.00,802.17)(23.700,-19.000){2}{\rule{0.313pt}{0.400pt}}
\multiput(658.00,782.92)(0.933,-0.493){23}{\rule{0.838pt}{0.119pt}}
\multiput(658.00,783.17)(22.260,-13.000){2}{\rule{0.419pt}{0.400pt}}
\multiput(682.00,771.58)(0.520,0.496){43}{\rule{0.517pt}{0.120pt}}
\multiput(682.00,770.17)(22.926,23.000){2}{\rule{0.259pt}{0.400pt}}
\multiput(706.00,792.92)(1.277,-0.491){17}{\rule{1.100pt}{0.118pt}}
\multiput(706.00,793.17)(22.717,-10.000){2}{\rule{0.550pt}{0.400pt}}
\multiput(731.00,784.58)(1.109,0.492){19}{\rule{0.973pt}{0.118pt}}
\multiput(731.00,783.17)(21.981,11.000){2}{\rule{0.486pt}{0.400pt}}
\multiput(755.00,793.94)(3.406,-0.468){5}{\rule{2.500pt}{0.113pt}}
\multiput(755.00,794.17)(18.811,-4.000){2}{\rule{1.250pt}{0.400pt}}
\multiput(779.00,789.93)(2.208,-0.482){9}{\rule{1.767pt}{0.116pt}}
\multiput(779.00,790.17)(21.333,-6.000){2}{\rule{0.883pt}{0.400pt}}
\put(804,783.17){\rule{4.900pt}{0.400pt}}
\multiput(804.00,784.17)(13.830,-2.000){2}{\rule{2.450pt}{0.400pt}}
\put(828,783.17){\rule{4.900pt}{0.400pt}}
\multiput(828.00,782.17)(13.830,2.000){2}{\rule{2.450pt}{0.400pt}}
\multiput(852.00,783.92)(0.659,-0.495){35}{\rule{0.626pt}{0.119pt}}
\multiput(852.00,784.17)(23.700,-19.000){2}{\rule{0.313pt}{0.400pt}}
\multiput(877.00,766.58)(1.225,0.491){17}{\rule{1.060pt}{0.118pt}}
\multiput(877.00,765.17)(21.800,10.000){2}{\rule{0.530pt}{0.400pt}}
\multiput(901.00,776.60)(3.406,0.468){5}{\rule{2.500pt}{0.113pt}}
\multiput(901.00,775.17)(18.811,4.000){2}{\rule{1.250pt}{0.400pt}}
\multiput(925.00,780.60)(3.552,0.468){5}{\rule{2.600pt}{0.113pt}}
\multiput(925.00,779.17)(19.604,4.000){2}{\rule{1.300pt}{0.400pt}}
\multiput(950.00,782.93)(1.550,-0.488){13}{\rule{1.300pt}{0.117pt}}
\multiput(950.00,783.17)(21.302,-8.000){2}{\rule{0.650pt}{0.400pt}}
\multiput(974.00,776.58)(0.600,0.496){37}{\rule{0.580pt}{0.119pt}}
\multiput(974.00,775.17)(22.796,20.000){2}{\rule{0.290pt}{0.400pt}}
\multiput(998.00,794.92)(0.595,-0.496){39}{\rule{0.576pt}{0.119pt}}
\multiput(998.00,795.17)(23.804,-21.000){2}{\rule{0.288pt}{0.400pt}}
\multiput(1023.00,773.94)(3.406,-0.468){5}{\rule{2.500pt}{0.113pt}}
\multiput(1023.00,774.17)(18.811,-4.000){2}{\rule{1.250pt}{0.400pt}}
\multiput(1047.00,771.58)(1.225,0.491){17}{\rule{1.060pt}{0.118pt}}
\multiput(1047.00,770.17)(21.800,10.000){2}{\rule{0.530pt}{0.400pt}}
\multiput(1071.00,779.93)(1.616,-0.488){13}{\rule{1.350pt}{0.117pt}}
\multiput(1071.00,780.17)(22.198,-8.000){2}{\rule{0.675pt}{0.400pt}}
\put(1096,772.67){\rule{5.782pt}{0.400pt}}
\multiput(1096.00,772.17)(12.000,1.000){2}{\rule{2.891pt}{0.400pt}}
\multiput(1120.00,772.93)(1.789,-0.485){11}{\rule{1.471pt}{0.117pt}}
\multiput(1120.00,773.17)(20.946,-7.000){2}{\rule{0.736pt}{0.400pt}}
\multiput(1144.00,765.92)(0.933,-0.493){23}{\rule{0.838pt}{0.119pt}}
\multiput(1144.00,766.17)(22.260,-13.000){2}{\rule{0.419pt}{0.400pt}}
\multiput(1168.00,754.58)(0.542,0.496){43}{\rule{0.535pt}{0.120pt}}
\multiput(1168.00,753.17)(23.890,23.000){2}{\rule{0.267pt}{0.400pt}}
\multiput(1193.00,775.94)(3.406,-0.468){5}{\rule{2.500pt}{0.113pt}}
\multiput(1193.00,776.17)(18.811,-4.000){2}{\rule{1.250pt}{0.400pt}}
\multiput(1217.58,773.00)(0.496,0.646){45}{\rule{0.120pt}{0.617pt}}
\multiput(1216.17,773.00)(24.000,29.720){2}{\rule{0.400pt}{0.308pt}}
\multiput(1241.58,801.73)(0.497,-0.559){47}{\rule{0.120pt}{0.548pt}}
\multiput(1240.17,802.86)(25.000,-26.863){2}{\rule{0.400pt}{0.274pt}}
\multiput(1266.00,776.59)(1.368,0.489){15}{\rule{1.167pt}{0.118pt}}
\multiput(1266.00,775.17)(21.579,9.000){2}{\rule{0.583pt}{0.400pt}}
\multiput(1290.00,783.93)(2.602,-0.477){7}{\rule{2.020pt}{0.115pt}}
\multiput(1290.00,784.17)(19.807,-5.000){2}{\rule{1.010pt}{0.400pt}}
\multiput(1314.00,778.93)(1.427,-0.489){15}{\rule{1.211pt}{0.118pt}}
\multiput(1314.00,779.17)(22.486,-9.000){2}{\rule{0.606pt}{0.400pt}}
\multiput(1339.58,771.00)(0.496,0.519){45}{\rule{0.120pt}{0.517pt}}
\multiput(1338.17,771.00)(24.000,23.928){2}{\rule{0.400pt}{0.258pt}}
\put(1363,795.67){\rule{5.782pt}{0.400pt}}
\multiput(1363.00,795.17)(12.000,1.000){2}{\rule{2.891pt}{0.400pt}}
\multiput(1387.00,795.92)(0.972,-0.493){23}{\rule{0.869pt}{0.119pt}}
\multiput(1387.00,796.17)(23.196,-13.000){2}{\rule{0.435pt}{0.400pt}}
\multiput(1412.00,782.93)(2.602,-0.477){7}{\rule{2.020pt}{0.115pt}}
\multiput(1412.00,783.17)(19.807,-5.000){2}{\rule{1.010pt}{0.400pt}}
\sbox{\plotpoint}{\rule[-0.500pt]{1.000pt}{1.000pt}}%
\put(1306,667){\makebox(0,0)[r]{random field}}
\multiput(1328,667)(20.756,0.000){4}{\usebox{\plotpoint}}
\put(1394,767){\usebox{\plotpoint}}
\put(220,874){\usebox{\plotpoint}}
\multiput(220,874)(17.601,-11.000){2}{\usebox{\plotpoint}}
\put(256.26,854.09){\usebox{\plotpoint}}
\multiput(269,849)(17.270,-11.513){2}{\usebox{\plotpoint}}
\put(312.44,829.76){\usebox{\plotpoint}}
\put(331.07,821.12){\usebox{\plotpoint}}
\put(349.11,819.15){\usebox{\plotpoint}}
\multiput(366,829)(20.595,2.574){2}{\usebox{\plotpoint}}
\put(408.40,829.79){\usebox{\plotpoint}}
\put(428.68,832.42){\usebox{\plotpoint}}
\put(448.82,832.55){\usebox{\plotpoint}}
\put(469.09,829.73){\usebox{\plotpoint}}
\multiput(488,832)(19.159,-7.983){2}{\usebox{\plotpoint}}
\put(529.16,823.43){\usebox{\plotpoint}}
\put(549.34,827.89){\usebox{\plotpoint}}
\put(569.19,828.06){\usebox{\plotpoint}}
\multiput(585,823)(19.690,6.563){2}{\usebox{\plotpoint}}
\put(626.91,821.30){\usebox{\plotpoint}}
\put(645.27,811.62){\usebox{\plotpoint}}
\multiput(658,805)(17.270,11.513){2}{\usebox{\plotpoint}}
\put(699.34,828.95){\usebox{\plotpoint}}
\put(718.93,828.38){\usebox{\plotpoint}}
\multiput(731,825)(14.985,14.361){2}{\usebox{\plotpoint}}
\put(767.93,837.76){\usebox{\plotpoint}}
\put(785.59,829.79){\usebox{\plotpoint}}
\multiput(804,832)(19.925,5.812){2}{\usebox{\plotpoint}}
\put(845.29,831.80){\usebox{\plotpoint}}
\put(865.31,826.87){\usebox{\plotpoint}}
\put(885.56,827.50){\usebox{\plotpoint}}
\put(905.58,831.05){\usebox{\plotpoint}}
\multiput(925,827)(19.987,5.596){2}{\usebox{\plotpoint}}
\put(966.12,837.36){\usebox{\plotpoint}}
\put(985.73,834.11){\usebox{\plotpoint}}
\put(1005.32,830.46){\usebox{\plotpoint}}
\multiput(1023,834)(20.595,2.574){2}{\usebox{\plotpoint}}
\put(1065.77,829.96){\usebox{\plotpoint}}
\put(1085.87,830.97){\usebox{\plotpoint}}
\put(1104.84,827.47){\usebox{\plotpoint}}
\multiput(1120,818)(18.250,9.885){2}{\usebox{\plotpoint}}
\put(1160.74,834.49){\usebox{\plotpoint}}
\put(1180.55,831.48){\usebox{\plotpoint}}
\multiput(1193,827)(15.620,13.668){2}{\usebox{\plotpoint}}
\put(1228.87,835.63){\usebox{\plotpoint}}
\multiput(1241,823)(20.182,-4.844){2}{\usebox{\plotpoint}}
\put(1279.19,830.74){\usebox{\plotpoint}}
\multiput(1290,842)(18.564,-9.282){2}{\usebox{\plotpoint}}
\put(1332.65,823.29){\usebox{\plotpoint}}
\put(1349.11,830.69){\usebox{\plotpoint}}
\multiput(1363,844)(20.756,0.000){2}{\usebox{\plotpoint}}
\put(1406.02,843.24){\usebox{\plotpoint}}
\put(1426.76,843.61){\usebox{\plotpoint}}
\put(1436,844){\usebox{\plotpoint}}
\end{picture}